	\documentstyle[12pt,group21]{article}
\title{Application of a symmetry adapted algebraic model to the
 vibrational spectrum of methane}
\author{R. Lemus\adr{1} F. P\'erez-Bernal\adr{2}
A. Frank\adr{1,3}  
R. Bijker\adr{1} \and \ J.M. Arias\adr{2}}

\address[1]{Instituto de Ciencias Nucleares, U.N.A.M.,\\
         A.P. 70-543, 04510 M\'exico D.F., M\'exico}
\address[2]{Departamento de F\'{\i}sica At\'omica, Molecular y Nuclear,\\
         Facultad de F\'{\i}sica, Universidad de Sevilla,\\
         Apdo. 1065, 41080 Sevilla, Espa\~na}
\address[3]{Instituto de F\'{\i}sica, Laboratorio de Cuernavaca,\\
         A.P. 139-B, Cuernavaca, Morelos, M\'exico}

\newcommand{\ba}{\begin{eqnarray}}
\newcommand{\ea}{\end{eqnarray}}
\def\ii{\'\i}
\begin{document}
\maketitle

\section{Algebraic model}
Recently, a symmetry-adapted version of the vibron 
model has been developed [1-4].
 The aim of this work  is to present the description of all
stretching and bending  
vibrations of the methane molecule   
in the framework of such a  model. 

In the present algebraic approach a $U(2)$ algebra is associated with 
each relevant interatomic interaction. For the CH$_4$ molecule we have 
four $U(2)$ algebras corresponding to the C-H interactions and six more 
representing the H-H couplings. The assignments and the choice of the 
Cartesian coordinate system are the same as in \cite{cinco}.
The molecular dynamical group is then given by the product $U_1 (2)
\otimes \ldots \otimes U_{10}(2)$,  where 
each $U_i(2)$ algebra ($i=1,\ldots,10$) is generated by the set 
$\{ \hat G_i \} \equiv \{ \hat N_i, \, \hat J_{+,i}, 
\, \hat J_{-,i}, \, \hat J_{0,i} \}$, which satisfies the commutation 
relations 
\ba
\, [ \hat J_{0,i}, \hat J_{\pm,i}] \;=\; \pm \hat J_{\pm,i} ~,
\hspace{1cm} 
\, [ \hat J_{+,i}, \hat J_{-,i}] \;=\; 2 \hat J_{0,i} ~,
\hspace{1cm} 
\, [ \hat N_i, \hat J_{\mu,i}] \;=\; 0 ~, \label{jmui}
\ea
with $\mu=\pm,0$. The labeling is such that $i=1,\ldots,4$ correspond to 
the C-H couplings while the other values of $i$ are associated with 
H-H interactions \cite{cinco}.
Here $\hat N_i$ is the $i$-th boson number operator.  
 All physical operators are 
expressed in terms of the generators $\{ \hat G_i \}$, and hence 
commute with the number operators $\hat N_i$. 
Since $\vec{J}^2_i=\hat N_i(\hat N_i+2)/4$ we can make the identification 
$j_i=N_i/2$. The eigenvalues of $\hat J_{0,i}$ are restricted to 
$m_i \geq 0$ and can have the values 
$m_i=N_i/2, (N_i-2)/2, \ldots , 1/2$ or 0 for $N_i$ odd or even, 
respectively \cite{seis}. The local basis states for each 
oscillator are usually written as $|N_i,v_i \rangle$, where 
$v_i=(N_i-2m_i)/2=0,1, \ldots [N_i/2]$ denotes the number of oscillator 
quanta in the $i$-th oscillator. 
For the CH$_4$ molecule there are two different boson 
numbers, $N_s$ for the C-H couplings and $N_b$ for the H-H couplings,
which correspond to the stretching and bending modes, respectively.

The  symmetry of methane is taken into account by projecting 
the local operators $\{ \hat G_i \}$, which act on bond $i$, on the 
irreducible representations $\Gamma$ of the tetrahedral group 
${\cal T}_d$. For the $\hat J_{\mu,i}$ generators of Eq.~(\ref{jmui})
we obtain the ${\cal T}_d$ tensors 
\ba
\hat T^{\Gamma_x}_{\mu,\gamma} &=& 
\sum_{i=1}^{10} \, \alpha^{\Gamma_x}_{\gamma,i} \, \hat J_{\mu,i} ~,
\label{alpha}
\ea
where  $\gamma$ denotes the component of $\Gamma$, 
and the label $x$ refers to stretching ($s$) or bending ($b$). 
The explicit expressions for the tensors  are  similar to the one
phonon wave functions expansions given in \cite{cinco}.  
The algebraic Hamiltonian is constructed by repeated couplings 
of these tensors to a total symmetry $A_1$.

The methane molecule has nine vibrational degrees of freedom. Four of 
them correspond to the fundamental stretching modes ($A_1 \oplus F_2$) 
and the other five to the fundamental bending modes ($E \oplus F_2$) 
\cite{siete}. The projected tensors of Eq.~(2) 
 correspond to ten degrees of freedom, four of which 
($A_1 \oplus F_2$) are related to stretching modes and six 
($A_1 \oplus E \oplus F_2$) to the bendings. Consequently we can 
identify the tensor $\hat{T}^{A_{1,b}}_{\mu,1}$ as the 
operator associated to a spurious mode. 
This identification makes it possible to eliminate the spurious states 
{\em exactly}. This can be achieved by (i) constructing the physical 
Hamiltonian by simply ignoring the $\hat{T}^{A_{1,b}}_{\mu,1}$ tensor 
in the coupling procedure, and (ii) diagonalizing this Hamiltonian 
in a symmetry-adapted basis from which the spurious mode 
has been removed following the procedure of \cite{cinco}.

According to the above procedure, we  construct the ${\cal T}_d$ 
invariant interactions that are at most quadratic in the generators 
and conserve the total number of quanta 
\ba
\hat{\cal H}_{\Gamma_x} &=& \frac{1}{2N_{x}} \sum_{\gamma} \left( 
  \hat T^{\Gamma_x}_{-,\gamma} \, \hat T^{\Gamma_x}_{+,\gamma}
+ \hat T^{\Gamma_x}_{+,\gamma} \, \hat T^{\Gamma_x}_{-,\gamma} 
\right) ~,
\nonumber\\
\hat{\cal V}_{\Gamma_x} &=& \frac{1}{N_{x}} \sum_{\gamma} \,
\hat T^{\Gamma_x}_{0,\gamma} \, \hat T^{\Gamma_x}_{0,\gamma} ~.
\label{hv}
\ea
Here $\Gamma=A_1$, $F_2$ for the stretching vibrations $x=s$ and 
$\Gamma=E$, $F_2$ for the bending vibrations $x=b$. In addition to
Eq.~(\ref{hv}), there are two stretching-bending interactions which
will not be considered at the moment. 

The zeroth order vibrational Hamiltonian is then written as
\ba
\hat H_0 &=& \omega_1 \, \hat{\cal H}_{A_{1,s}} 
   + \omega_2 \, \hat{\cal H}_{E_b} 
   + \omega_3 \, \hat{\cal H}_{F_{2,s}} 
   + \omega_4 \, \hat{\cal H}_{F_{2,b}} 
\nonumber\\
&& + \alpha_2 \, \hat{\cal V}_{E_b}  
   + \alpha_3 \, \hat{\cal V}_{F_{2,s}} 
   + \alpha_4 \, \hat{\cal V}_{F_{2,b}} ~. \label{h0}
\ea
The interaction $\hat{\cal V}_{A_{1,s}}$ has not been included since 
the combination $\sum_\Gamma (\hat{\cal H}_{\Gamma_s} + \hat{\cal
V}_{\Gamma_s})$ corresponds to the constant $N_s +2$. A similar
 situation is present  for the  
bending interactions, but in this case the interaction 
$\hat{\cal V}_{A_{1,b}}$ has already been excluded in order to remove 
the spurious $A_1$ bending mode. The subscripts of the parameters 
correspond to the $(\nu_1,\nu_2^{l_2},\nu_3^{l_3},\nu_4^{l_4})$ 
labeling of a set of basis states for 
the vibrational levels of CH$_4$. Here $\nu_1$, $\nu_2$, $\nu_3$ and 
$\nu_4$ denote the number of quanta in the $A_{1,s}$, $E_b$, 
$F_{2,s}$ and $F_{2,b}$ modes, respectively. The labels $l_i$ are 
related to the vibrational angular momentum associated with degenerate 
vibrations \cite{siete}.
  In the harmonic limit the interactions of Eqs.~(\ref{hv}) 
and~(4) attain a particularly simple form, which can be 
directly related to configuration space interactions [1-4]. 
The $\hat{\cal H}_{\Gamma_x}$ terms represent the 
anharmonic counterpart of the harmonic interactions, while 
the $\hat{\cal V}_{\Gamma_x}$ terms are purely anharmonic contributions 
which vanish in the harmonic limit. 

The zeroth order Hamiltonian of Eq.~(\ref{h0}) is not sufficient to 
obtain a high-quality fit of the vibrations of methane. Several other 
physically meaningful 
interaction  are essential for such a fit.   
   For the study of the 
vibrational excitations of methane we use the ${\cal T}_d$ invariant 
Hamiltonian
\ba
\hat H &=& \omega_1 \, \hat{\cal H}_{A_{1,s}} 
         + \omega_2 \, \hat{\cal H}_{E_b} 
         + \omega_3 \, \hat{\cal H}_{F_{2,s}} 
         + \omega_4 \, \hat{\cal H}_{F_{2,b}} 
         + \alpha_3 \, \hat{\cal V}_{F_{2,s}} 
\nonumber\\
&& + X_{11} \left( \hat{\cal H}_{A_{1,s}} \right)^2
   + X_{22} \left( \hat{\cal H}_{E_{b}  } \right)^2
   + X_{33} \left( \hat{\cal H}_{F_{2,s}} \right)^2
   + X_{44} \left( \hat{\cal H}_{F_{2,b}} \right)^2
\nonumber\\
&& + X_{12} \left( \hat{\cal H}_{A_{1,s}} \, \hat{\cal H}_{E_b    } \right)
   + X_{14} \left( \hat{\cal H}_{A_{1,s}} \, \hat{\cal H}_{F_{2,b}} \right)
\nonumber\\
&& + X_{23} \left( \hat{\cal H}_{E_b    } \, \hat{\cal H}_{F_{2,s}} \right) 
   + X_{24} \left( \hat{\cal H}_{E_b    } \, \hat{\cal H}_{F_{2,b}} \right) 
   + X_{34} \left( \hat{\cal H}_{F_{2,s}} \, \hat{\cal H}_{F_{2,b}} \right) 
\nonumber\\
&& + g_{22} \, \left( \hat l^{A_2} \right)^2  
   + g_{33} \, \sum_{\gamma} \hat l^{F_1}_{s,\gamma} \, 
                             \hat l^{F_1}_{s,\gamma} 
   + g_{44} \, \sum_{\gamma} \hat l^{F_1}_{b,\gamma} \, 
                             \hat l^{F_1}_{b,\gamma} 
   + g_{34} \, \sum_{\gamma} \hat l^{F_1}_{s,\gamma} \, 
                             \hat l^{F_1}_{b,\gamma} 
\nonumber\\
&& + t_{33} \, \hat{\cal O}_{ss}
   + t_{44} \, \hat{\cal O}_{bb}
   + t_{34} \, \hat{\cal O}_{sb}
   + t_{23} \, \hat{\cal O}_{2s}
   + t_{24} \, \hat{\cal O}_{2b} ~. \label{hamilt}
\ea

The $X_{ij}$ terms are quadratic in the operators 
$\hat{\cal H}_{\Gamma_x}$ and hence represent anharmonic vibrational 
interactions. The $g_{ij}$ terms are related to the 
vibrational angular momenta associated with the degenerate vibrations
 [1-3]. 

  In the harmonic limit the expectation   
value of the diagonal terms in Eq.~(\ref{hamilt}) leads to the familiar 
Dunham expansion \cite{siete}. 
\ba
\sum_i \omega_i \, (v_i + \frac{d_i}{2}) + \sum_{j \geq i} \sum_i 
X_{ij} \, (v_i + \frac{d_i}{2}) (v_j + \frac{d_j}{2})
+ \sum_{j \geq i} \sum_i g_{ij} \, l_i l_j ~. \label{Dunham}
\ea
Here $d_i$ is the degeneracy of the vibration. 
 The $t_{ij}$ terms in Eq.~(\ref{hamilt})
give rise to further splittings of the vibrational levels 
$(\nu_1,\nu_2,\nu_3,\nu_4)$ into its possible sublevels \cite{diez}. 
  Their explicit expressions are given in ref. [2].

\section{Results}  
The Hamiltonian of Eq.~(\ref{hamilt}) involves 23 interaction 
strengths and the two boson numbers, $N_s$ and $N_b$. The vibron 
number associated with the stretching vibrations is determined 
from the spectroscopic constants $\omega_e$ and $x_e \omega_e$ 
for the CH molecule to be $N_s=43$ \cite{once}. The vibron number 
for the bending vibrations, which are far more harmonic than the 
stretching vibrations, is taken to be $N_b=150$. We have carried out 
a least-square fit to the vibrational spectrum of methane including 
44 energies.  We find an overall fit to the observed levels 
with a r.m.s. deviation which is an order of magnitude better than 
in previous studies. While the r.m.s. deviations of \cite{cinco} and 
\cite{doce} are 12.16 and 11.61 cm$^{-1}$ for 19 energies, we find a 
r.m.s. of 1.16 cm$^{-1}$ for 44 energies.  The values of the fitted
parameters as well as all predicted levels up to $V = 3$ are given in
ref. [1].

The $\alpha_3$ term is completely anharmonic in origin and has no 
counterpart in the harmonic limit. 
In order to address the importance of this term in 
Eq.~(\ref{hamilt}) we have carried out another calculation without 
this term. With one less interaction term the r.m.s. deviation 
increases from 1.16 to  4.48 cm$^{-1}$. This shows the importance of 
the term proportional to $\alpha_3$ to obtain an accurate description 
of the anharmonicities that are present in the data. 
 The  absence of the $\alpha_3$ term in the second calculation  can 
only partially be compensated by  the anharmonicity 
constants $X_{ij}$ \cite{uno}.

\section{Conclusions}
In summary, in this work we present the description of vibrational
excitations  
of methane in a symmetry-adapted algebraic model. We find an overall fit 
to the 44 observed levels with a r.m.s. deviation of 1.16 cm$^{-1}$, 
which can be considered of spectroscopic quality. 
We pointed out that the ${\cal V}_{F_{2,s}}$ term in combination with 
the anharmonic effects in the other interation terms plays a 
crucial role in obtaining a fit of this quality. Purely anharmonic terms 
of this sort arise naturally in the symmetry-adapted algebraic model, 
but vanish in the harmonic limit. Physically, these contributions arise
from the anharmonic character of the interatomic interactions, and seem 
to play an important role when dealing with molecular 
anharmonicities.

These studies suggest that the symmetry-adapted algebraic model 
provides a numerically efficient tool to study molecular 
vibrations with high precision. The main difference with other
algebraic   
methods is the use of symmetry-adapted tensors in the construction 
of the Hamiltonian. In this approach, the interactions can be constructed 
in a systematic way, each term has a direct physical interpretation, 
and spurious modes can be eliminated exactly.

\section*{Acknowledgments}
We thank Prof. J.C. Hilico for his interest and for making available 
to us his compilation of observed level energies. 
This work was supported in part by the 
European Community under contract nr. CI1$^{\ast}$-CT94-0072, 
DGAPA-UNAM under project IN105194, CONACyT-M\'exico under project 
400340-5-3401E and Spanish DGCYT under project PB92-0663.

\end{document}